\documentclass[10pt]{article}  
\newcommand{\comma}{\;\; ,}
\newcommand{\period}{\;\; .}

\newcommand{\eq}{\; = \;}
\newcommand{\sep}{\;\; , \;\;}
\newcommand{\be}{\begin{equation}}
\newcommand{\bd}{\begin{displaymath}}
\newcommand{\ee}{\end{equation}}
\newcommand{\ed}{\end{displaymath}}
\newcommand{\ba}{\begin{eqnarray}}
\newcommand{\ea}{\end{eqnarray}}

\newcommand{\integ}[1]{\textstyle [\frac{#1}{N}] }

\newcommand{\minus}{\! - \! }

\title{The Riemann surface of the chiral Potts model free energy function}

\author{ R.J. Baxter\\
{\protect \small Theoretical Physics, I.A.S. and School of Mathematical
Sciences}\\
{\protect \small  The Australian National University,
 Canberra, A.C.T. 0200, Australia  }
\footnote{This work supported in part by the Australian Research Council}
}

\date{}

\begin{document}


\maketitle

\abstract{In a recent paper we derived the free energy or partition function of the $N$-state chiral Potts 
model by using the infinite lattice ``inversion relation'' method, together 
with a non-obvious extra symmetry. This gave us three recursion relations for 
the partition function per site $T_{pq}$  of the infinite lattice. Here we use these recursion 
relations to obtain the full Riemann surface of $T_{pq}$. In terms of the 
$t_p, t_q$ variables, it consists of an infinite number of Riemann sheets, each 
sheet corresponding to a point on a $(2N-1)$-dimensional lattice (for  $N > 2$).

The function $T_{pq}$ is  meromorphic  on this surface: we obtain 
the orders of all the zeros and poles. For $N$ odd, we show that these orders are determined by
the usual inversion and rotation relations (without the extra symmetry), together with a simple linearity ansatz.
For $N$ even, this method does not give the orders uniquely, but leaves only $[(N+4)/4]$ 
parameters to be determined.

}



\section{Introduction}

The free energy  of the $N$-state chiral Potts model was obtained implicitly in 1988 
\cite{Baxter1988} from 
the star-triangle relation, and explicitly in 1990 \cite{Seoul90}, \cite{Baxter90} from the 
functional relations 
for the finite-lattice transfer matrices. Recently \cite{paper1} we considered the 
infinite-lattice limit 
of these functional relations, and showed that they reduced to the rotation 
and inversion relations (which exist for most two-dimensional models, solvable or not 
\cite{Baxter82}), or rather to a single combined rotation-inversion relation. All other 
equations were 
either definitions of the auxiliary function $\tau_2 (t_q)$, or direct consequences of 
the relations and definitions, so by themselves contained no new information.

We showed that the reason we could solve the infinite relations was that we knew the 
analyticity properties in some ``central'' domain that included part of the physical regime. 
In particular, we knew that $\tau_2 (t_q)$ had only one branch cut in the $t_q$-plane, instead of 
the $N$ cuts one might expect. This was equivalent to an extra symmetry for the free energy. 
It ensured that one could use Wiener-Hopf methods to calculate the partition function per 
site $T_{pq}$.

We found that $T_{pq}$ satisfied three ``recursion'' relations: equations (70), (71) and (75)
of \cite{paper1}. (We could have obtained these directly from the solution found in 1990.)
Here we use these three relations to obtain the full Riemann surface 
(in both the $p$ and the $q$ rapidity variables)  on which $T_{pq}$ lives. It is a meromorphic 
function on this surface, its only singularities being poles. Its poles and zeros occur only when
$t_p^N = t_q^N$, $\lambda_p = \lambda_q^{\pm 1} $.

Regarding $T_{pq}$ as a function of $t_p, t_q$, its Riemann surface consists 
of an infinite number of Riemann sheets. Each sheet is specified by a set of $N$ integers 
$m_0, \ldots, m_{N-1}$ associated with the $p$-variable, and another set 
$n_0, \ldots, n_{N-1}$ associated with the $q$-variable. Thus it corresponds to a point on
a $2N$-dimensional lattice. In fact, although the chiral Potts model does not have the usual
difference property (that one can choose the rapidites $p,q$ so that the Boltzmann weights, 
correlations and free energy depend only on $p, q$ via their difference $p - q$), we do find a 
weak residual version of this property: $T_{pq}$ is the same on all sheets obtained by incrementing
$m_0, \ldots, m_{N-1}, n_0, \ldots, n_{N-1}$ by an arbitrary integer. Hence the 
Riemann surface can be associated with
a lattice of reduced dimension $2N-1$.

The case $N=2$, when the model reduces to the Ising model, is special. Then two of the integers
$m_0, \ldots, n_{N-1}$ vanish, so the surface is one-dimensional, being that of the 
elliptic function  argument $z = \exp[\pi (u_p - u_q)/K']$ of equation (B7) of \cite{paper1}.

A knowledge of the Riemann surface, and of the orders of the poles and zeros, gives us a great deal of information
about the function $T_{pq}$. We observe that that these orders are linear in the 
$m_0, \ldots, m_{N-1}$, and in the $n_0, \ldots, n_{N-1}$. Hence they are bilinear in the full set
of $2N$ integers $m_0, \ldots, m_{N-1}, n_0, \ldots, n_{N-1}$, depending only on their 
differences.

We ask the question: is this bilinearity, together with the basic inversion and rotation relations
sufficient to determine the orders of the poles and zeros? We find that for $N$ odd the 
answer is yes.
For $N$ even it is not quite sufficient: there are $[(N+4)/4]$ parameters still undetermined.

\section{Domains and Riemann sheets}

The Boltzmann weights of the $N$-state chiral Potts model are functions of a 
constant $k$,  and two ``rapidity''variables
$p$ and $q$.\cite{Seoul90, Baxter90} Let 
\be  k' = (1-k^2)^{1/2} \sep \omega = e^{2 \pi i /N} \period \ee
Then the variable $p$ can be thought of as a point 
$(x_p,y_p,t_p,\lambda_p)$ 
on the algebraic curve 
\bd x_p^N+y_p^N = k (1+x_p^N y_p^N) \sep t_p = x_p y_p \comma \ed
\be \label{curve} k \, x_p^N  = 1-k'/\lambda_p \sep k \, y_p^N  = 1-k'\lambda_p \period \ee
Similarly $q$ is the point $(x_q,y_q,t_q)$.

As we remark in \cite{paper1}, if $x_p, x_q,y_p,y_q,\omega x_p $ all lie on the unit circle 
and are ordered anti-cyclically around it, then all the Boltzmann weights are real and 
positive, and therefore so is the partition function per site $T_{pq}$. We refer to this case
 as the {\em physical regime}. Outside this regime we define $T_{pq}$ by analytic continuation.

This function $T_{pq}$ therefore lives on a Riemann surface in both the $p$ and 
the $q$ variables.
To specify this surface, first consider the $p$ variable. If $|\lambda_p| < 1$, then
$x_p$ lies in the region $\cal S$ of Figure 1, while $y_p$ lies in one of the 
$N$ approximately circular regions
${\cal R}_0, \ldots , {\cal R}_{N-1}$ surrounding the points 
$1, \omega, \ldots, \omega^{N-1}$. 

These regions shrink to points in the 
low-temperature limit $k' \rightarrow 0$, so in this limit it is certainly true that 
\bd y_p \simeq \omega^r  \; \; \; {\rm if} \; \; \; y_p \in {\cal R}_r \period \ed
We find to helpful to write ``$\; y_p \simeq \omega^r \; $'', by which we mean
 ``$\; y_p \in {\cal R}_r\; $''.

We define a ``domain'' ${\cal D}_r$
for $x_p$ and $y_p$ simultaneously 
by saying that if $y_p$ lies in ${\cal R}_r$ , then
$(x_p,y_p)$ lies in the domain ${\cal D}_r$. More simply, we say that 
$p$ lies in ${\cal D}_r$. We also say in this case that $p$ has {\em  parity} 0
and that it has {\em type} $r$.

Conversely, if $|\lambda_p| > 1$, then
$y_p$ lies in the region $\cal S$  and $x_p$ lies in one of
${\cal R}_0, \ldots , {\cal R}_{N-1}$. If  $x_p$ lies in ${\cal R}_r$ , then
we say that $(x_p,y_p)$ or $p$ lies in the domain ${\cal D}'_r$,   has parity 1 and type $r$.

Let $\rho, r$ be the parity and type  of $p$. Then
\ba
x_p \in {\cal S} \sep  y_p \in {\cal R}_r  \sep |\lambda_p| < 1 \! \! \! & {\rm and} & p \in {\cal D}_r \; \; \; \; {\rm if} \; \;  \rho = 0 \comma \nonumber \\
 x_p \in {\cal R}_r \sep y_p \in {\cal S} \sep |\lambda_p| > 1  \! \!& {\rm and} &   p \in {\cal D}'_r \; \; \; \;  {\rm if} \; \;  \rho = 1 \period  \ea
 
We refer to the case when $p$ and $q$ both lie in ${\cal D}_0$, so that  $|\lambda_p| < 1 $, 
$|\lambda_q| < 1$, $y_p, y_q \in {\cal R}_0$,  
$x_p, x_q \in {\cal S}$, as the {\em central} regime or domain. It overlaps the physical regime, 
so  $T_{pq}$ is readily extended to this domain. Series expansions strongly suggest that it 
has no poles or zeros in the central domain (as can be verified from the  explicit solution 
(67) of \cite{paper1}).

If we analytically continue from the central domain 
to $|\lambda_p| > 1$, then as $\lambda_p$ crosses the unit circle, $x_p$ enters one 
of the regions ${\cal R}_0, \ldots , {\cal R}_{N-1}$, say ${\cal R}_{r_1}$, while
$y_p$ leaves ${\cal R}_0$ and enters $\cal S$. Thus $p$ goes from the domain
${\cal D}_0$ to ${\cal D}'_{r_1}$. Its type changes from $0$ to $r_1$, and its parity
from 0 to 1.

We can then analytically continue from $|\lambda_p| > 1$ back to $|\lambda_p| < 1$,
but $p$ will not necessarily return to ${\cal D}_0$: in general it will go to a new domain
${\cal D}_{r_2}$. And so on: $p$ will move successively through a sequence of domains
\be \label{seq}
{\cal D}_0 \; , \; {\cal D}'_{r_1} \; , \; {\cal D}_{r_2} \; , \; {\cal D}'_{r_3}
 \; , \; {\cal D}_{r_4} \; , \ldots  \ee
Their types are $0, r_1, r_2, r_3, r_4, \ldots$, and their parities are $0,1,0,1,0, \ldots$.

From (\ref{curve}),
\bd k^2 \, t_p^N \eq 1 - k' (\lambda_p + 1/\lambda_p) + {k'}^2 \comma \ed
so the unit circle in the $\lambda_p$-plane corresponds to $N$ straight-line segments
in the $t_p$-plane, from $\omega^j \eta$ to $\omega^j /\eta$, as in Figure 2 of \cite{paper1}. Here
$\eta = [(1-k')/(1+k')]^{1/N}$. In terms of the $t_p$ variable, each $(x_p, y_p)$ domain
is therefore a Riemann sheet, with branch cuts along these straight-line segments. As 
one goes from a domain to a neighbouring domain, $t_p$ crosses one of these $N$ branch cuts and goes 
from one sheet to the next. We shall use the words ``domain'', ``Riemann sheet'' and ``sheet''
 interchangeably.

We define $T_{pq}$ by analytic continuation as $p$ moves through this sequence, starting from 
some initial value in ${\cal D}_0$. As $p$ moves from one domain to the next,
$T_{pq}$ moves from one Riemann sheet to the next.

At this stage we do not know the Riemann surface on which 
$T_{pq}$ lives (i.e. how the Riemann sheets connect with one another), 
so we must be prepared to remember this full 
sequence of domains in order to uniquely specify 
the value of $T_{pq}$ at a point $p$. Since the parities alternate, it is sufficient
to remember the domain types  $\{r_1, r_2, r_3, r_4, \ldots \}$.

Similarly, if initially the other rapidity variable $q$ is in the central domain ${\cal D}_0$, 
and is analytically continued
through the sequence
\bd
{\cal D}_0 \; , \; {\cal D}'_{s_1} \; , \; {\cal D}_{s_2} \; , \; {\cal D}'_{s_3}
 \; , \; {\cal D}_{s_4} \; , \ldots  \ed
we must remember
the sequence $\{s_1, s_2, s_3, s_4, \ldots \}$.

Fortunately it seems that  $T_{pq}$ does not depend on whether $p$ or $q$ moves first:
i.e. how the two sequences $\{r_1, r_2, r_3, r_4, \ldots \}$, $\{s_1, s_2, s_3, s_4, \ldots \}$
are interleaved. Thus the Riemann surface in both variables is simply the union of 
the two single-variable surfaces.

\begin{figure}[hbt]
\begin{picture}(420,274) (-40,-74)
\put(70,100) {\line (1,0) {200}}
\put(170,15) {\line (0,1) {170}}
\put(240,100){\circle{60}}
\put(135,161){\circle{60}}
\put(135,39){\circle{60}}
\put(237,97){$\bullet$}
\put(132,158){$\bullet$}
\put(132,36){$\bullet$}
\put(158,88){O}
\put(212,150){${\cal S}$}
\put(230,87){${\cal R}_0$}
\put(126,149){${\cal R}_1$}
\put(121,29){${\cal R}_{N-1}$}
\put(0,-15) {Figure 1: The $N+1$ regions ${\cal S}, {\cal R}_0, \ldots , {\cal R}_{N-1}$ of the complex plane  }
\put(35,-32){in which $x_q$ and $y_q$ lie  (for $N = 3$). ${\cal R}_0, \ldots {\cal R}_{N-1}$ are the}
\put(35,-49){interiors of the approximate circles centred on $1, \omega , \ldots ,\omega^{N-1}$.}
\put(35,-66){ ${\cal S}$ is the the complex plane outside all $N$ circles.}
\end{picture}
\label{xyplane}
\end{figure}

\section{Recursion relations for $T_{pq}$}

In \cite{paper1} we obtained the two relations, true for 
 $r, s = 0, \ldots , N-1$:
\ba \label{Trcontp}
T_r(\omega^r y_p, x_p|x_q,y_q ) & =  & \frac{N \, T(\omega^r x_p,y_p|x_q,y_q )}
{T(\omega^{-1} x_p,y_p|x_q,y_q ) \, T( x_p,y_p|x_q,y_q )} 
\; \prod_{j=1}^r \frac{t_p - \omega^{-j} t_q}{
(x_p \!  -  \!  \omega^{-j} x_q) (y_p  \! -  \! \omega^{-j} y_q) } \; \nonumber \\ 
& & \times \prod_{j=r+1}^{N-1} \frac{t_p - \omega^{-j} t_q}{
(x_p - \omega^{-j} y_q) (y_p - \omega^{-j} x_q) }
 \comma \ea
\ba \label{Trcontq}
T_s(x_p,y_p|\omega^s y_q, x_q) & =  & \frac{N \, T(x_p,y_p|\omega^s x_q,y_q)}
{T(x_p,y_p|\omega^{-1} x_q,y_q) T(x_p,y_p| x_q,y_q)} \prod_{j=1}^s \frac{t_p - \omega^{j-1} t_q}{
(x_p \!  -  \!  \omega^{j-1} y_q) (y_p  \! -  \! \omega^j x_q) } \; \nonumber \\ 
& & \times \prod_{j=s+1}^{N-1} \frac{t_p - \omega^{j-1} t_q}{
(x_p - \omega^{j-1} x_q) (y_p - \omega^j y_q) }
 \comma \ea
together with 
\be \label{2ndratio}
\frac{T(x_p, y_p| x_q, y_q) \, T(\omega x_p, y_p| \omega x_q, y_q)}{
T(x_p, y_p|\omega  x_q, y_q) \, T(\omega x_p, y_p|  x_q, y_q)} \eq 
\frac{(x_p - x_q)(\omega t_p - t_q)}{(\omega x_p - x_q)( t_p - t_q)} \period \ee

In these three relations the $p$-arguments $(\omega^i x_p,y_p)$ (for any $i$) of the function $T$ 
(without a suffix) lie 
in the initial domain ${\cal D}_0$, and so do the $q$-arguments $(\omega^ix_q,y_q)$.
 The function $T_r$ on the lhs of (\ref{Trcontp}) has $p$-arguments $(x'_p, y'_p) =
 (\omega^r y_p, x_p)$
lying in the domain ${\cal D}'_r$ adjacent to ${\cal D}_0$. Similarly, 
$T_s$ on the lhs of (\ref{Trcontq}) has $q$-arguments $(x'_q, y'_q) =
 (\omega^s y_q, x_q)$
lying in the domain ${\cal D}'_s$ adjacent to ${\cal D}_0$.

We can use  (\ref{Trcontp}) and  (\ref{Trcontq}) to obtain the function $T$ on any of its 
infinitely many Riemann sheeets. For instance, if in (\ref{Trcontp}) we allow 
$p = (x_p,y_p)$ to leave ${\cal D}_0$ and enter the neighbouring domain ${\cal D}'_m$,
where $x_p \simeq \omega^m$, then the arguments
of $T$ on the lhs will move to a sheet ${\cal D}_{rm}$. 
The three $T$s on the rhs move to the domains  
${\cal D}'_{m+r}$, ${\cal D}'_{m-1}$, ${\cal D}'_{m}$, respectively.
 These can in turn be expressed in terms of $T$ 
in ${\cal D}_0$ by again using (\ref{Trcontp}).

We can repeat this procedure {\em ad infinitem}. At each stage the function $T_{pq}$ 
on the left-hand side of (\ref{Trcontp}) moves to a new domain or sheet, while the ones on the rhs 
move to domains or sheets that have already been expressed in terms of the values of
$T_{pq}$ in the inital domain ${\cal D}_0$. Similarly, we can use (\ref{Trcontq})
to obtain $T_{pq}$ on successive sheets in the $q$-variable.

At first sight we will obtain an exponentially infinite Cayley tree, 
each $(p,q)$ surface having  $2N$ neighbours ($N$ neighbours in the $p$ variable, $N$ in 
the $q$-variable). The number of $m$th neighbours of the initial domain
(where $p, q \in {\cal D}_0$ ) will be
$2N \times (2N-1)^{m-1}$. 

However, it is not as as bad as that.
Consider the function $T_{pq}$ on a Riemann sheet of $p$-parity $\rho$ and $p$-type $r$,
and  of $q$-parity $\sigma$ and $q$-type $s$. Thus $\rho,\sigma = 0, 1$ and
$r,s = 0, \ldots , N \! - \! 1$.
Define an associated function $A_{rs}^{\rho \sigma} (a_p,b_p |a_q,b_q)$ by
\ba \label{defA}
T(x_p,y_p |x_q,y_q ) & = & A_{rs}^{00} (x_p,\omega^{-r} y_p |x_q,\omega^{-s} y_q)
 \; \; {\rm if} \; \; 
\rho = 0, \sigma = 0 \comma \nonumber \\
& = & 
N/A_{rs}^{01} (x_p,\omega^{-r} y_p |y_q,\omega^{-s} x_q) \; \; {\rm if} \; \; 
\rho = 0, \sigma = 1 \comma \nonumber \\
& = & 
N/A_{rs}^{10} (y_p,\omega^{-r} x_p |x_q,\omega^{-s} y_q)  \; \; {\rm if} \; \; 
\rho = 1, \sigma = 0 \comma \\
& = & A_{rs}^{11} (y_p,\omega^{-r} x_p |y_q,\omega^{-s} x_q) \; \; {\rm if} \; \; 
\rho = 1, \sigma = 1 \period \nonumber \ea
In each case the $p$-arguments $a_p,b_p$ of $A_{rs}^{\rho \sigma} (a_p,b_p |a_q,b_q)$
lie in ${\cal D}_0$, and so do the $q$-arguments $a_q, b_q$.

Then no matter how many times we iterate (\ref{Trcontp}) and (\ref{Trcontq}), the 
resulting function
$T_{pq}$ will have an associated function of the form
\be \label{Ars}
A_{rs}^{\rho \sigma} (a_p,b_p |a_q,b_q) \eq 
\prod_{j=0}^{N-1} [(a_p - \omega^j a_q)^{\alpha_j}
(a_p - \omega^j b_q)^{\alpha'_j} \ee  \bd \times (b_p - \omega^j a_q)^{\beta_j}
(b_p - \omega^j b_q)^{\beta'_j} (a_p b_p - \omega^j a_q b_q)^{\gamma_j} ] \times
\prod_{i=0}^{N-1} \prod_{j=0}^{N-1} T(\omega^i a_p, b_p| \omega^j a_q, 
b_q)^{\rho_{ij} } \comma \ed
where the exponents $ \alpha_j, \ldots , \rho_{ij}$ are integers and the functions 
$T$ on the rhs all have $p$ and $q$ arguments 
lying in the initial sheet ${\cal D}_0$. 
 
Further, we can 
always use (\ref{2ndratio}) to eliminate the $T(\omega^i a_p, b_p| 
\omega^j a_q,  b_q)$
with $i$ and $j$ greater than zero, so we can require that
\be 
 \rho_{ij} \eq 0 \; \; \; {\rm if } \; \; \; i \; \;  {\rm and} \; \;  j> 0  \comma \ee
leaving only $\rho_{00}$, $\rho_{i0}$ and $\rho_{0j}$ ($i,j >0$) as possibly non-zero.
There are then at most $ 5N + 2N-1 = 7N-1$ non-zero exponents. Equating any two Riemann 
surfaces with the same values of the function
$T_{pq}$,  it follows that any Riemann surface (in 
both the $p$ and $q$ variables) 
can be specified by at most $7N \! - \! 1$ arbitrary integers (together with the two-valued integers 
$\rho, \sigma$ and the $N$-valued integers $r,s$), so can be regarded as 
corresponding to a point on a $(7 N \! - \! 1)$-dimensional lattice.

In fact the lattice is of lower dimension still, since there are many relations 
between the exponents. For any rational number $x$, let $[x]$ denote its integer part,
so that
\bd [x] \leq x < [x]+1 \comma \ed
and define
\be F_{i,j} \eq \integ{i-1} - \integ{i-j-1}  - \integ{j} \period \ee
Then  $F_{ij}$ has value 0 or 1, and is periodic in $i,j$ of period $N$, i.e.   $F_{i+N,j} = F_{i,j+N} = F_{ij}$.

The relations between the exponents  are conveniently expressed by associating with 
each Riemann sheet  two 
more sets of integers:\footnote{We originally found
the $ n_j$ by keeping the $p$ variables in ${\cal D}_0$ and looking at the orders of 
the zeros of $T_{pq}$, considered as a function of $q$. Then $m_j = 0$ and the  rhs of (\ref{zeros})  becomes $2 n_{-k} + 
w_k$.
 In the normalization
where the Boltzmann weights
are free of poles we lose the $w_k$ and the zeros are simply $2 (-1)^{\sigma} \, n_{-k}$.
Similarly, one can obtain the $m_j$.}
\bd  m = \{  m_0, \ldots ,m_{N-1} \} \sep n = \{  n_0, \ldots ,n_{N-1} \} \period \ed
The set $m$ is varied when the $p$-domain is changed, and $n$ when the $q$-domain is changed.
If $m$ is the set on a sheet of parities $\rho, \sigma$ and types $ r, s$, and $m'$ is the set $m$ on 
a $p$-neighbouring sheet of parities $\rho', \sigma$ and types $ r', s$ 
(the $q$ variables being the same on each sheet), 
then $\rho' = 1-\rho$ and 
\be \label{mnbrs}
m'_j \eq m_{r+r'} + m_{r+r'+1} - m_j + \rho (1 - \delta_{r+r',0} ) - F_{j, r+r'} \comma \ee
for $j = 0, \ldots , N-1$, using the periodic convention
\be \label{periodic} m_{j+N} = m_j \sep \forall j\period \ee

Similarly, the set $n$ on a sheet of parities $\rho, \sigma$ and types $ r, s$ is 
related to the set $n'$ on a $q$-neighbouring sheet of parities $\rho, \sigma'$ and 
types $ r, s'$ by $\sigma' = 1-\sigma$ and
\be  \label{nnbrs}
n'_j \eq n_{s+s'} + n_{s+s'+1} - n_j + \sigma (1 - \delta_{s+s',0} ) - F_{j, s+s'} 
\period \ee

These recursion relations, together with the  conditions
\bd m_0 = \cdots m_{N-1} = n_0 \cdots = n_{N-1} = 0 \; \; \; {\rm in \;  the \;  initial \;  
sheet, \;  where \; } 
p,q \in {\cal D}_0 \ed
define the integers $m_0, \ldots, m_{N-1}, n_0, \ldots , n_{N-1}$ on all 
sheets. In particular, it follows that
\be \label{defmunu} \sum_{j=0}^{N-1} m_j = N \mu - r \sep \sum_{j=0}^{N-1} n_j = 
N \nu - s  \comma \ee
$\mu$ and $\nu$ being integers.

Define 
\be \label{defgint}
g_k \eq \sum_{i=0}^{N-1} m_{i+k} n_i \period \ee

Numerical Fortran experiments (in integer arithmetic) for small $N$ and sheets close to $\cal D$ strongly suggest that
\ba \label{reslts}
\alpha_j & = & \mu - \nu - m_{j+r} - m_{j+r+1} +  n_{s-j-1} +
 n_{s-j} + \tilde{\alpha}_j \comma \nonumber \\
\alpha'_j & = & -\mu - \nu + m_{j+r} + m_{j+r+1} + {\tilde{\alpha}}'_j \comma \nonumber 
\\
\beta_j & = & \mu + \nu - n_{s-j} - n_{s-j+1} + {\tilde{\beta}}_j \comma \nonumber \\
{\beta}'_j & = & -\mu + \nu  + {\tilde{\beta}}'_j \comma  \\
{\gamma}_j & = & 2 \, g_{j+r-s+1} - 2 \, g_{j+r-s-1}  -
 2 \, m_{j+r-s+\sigma} +  2 \, n_{s-r-j+\rho} + \nonumber \\
    & &     m_{j+r} + m_{j+r+1}   - n_{s-j-1} - 
n_{s-j} + {\tilde{\gamma}}_j   \comma \nonumber \\
{\rho}_{ij} & = & (m_{r-i+1} - m_{r-i-1})\delta_{j,0} +
(  n_{s-j+1} - n_{s-j-1})\delta_{i,0}  + {\tilde{\rho}}_{ij} \comma \nonumber\ea
where 
\ba \label{reslts1}
\tilde{\alpha}_j & = & \integ{j} + \integ{j-\rho(1-\sigma)} - 
\integ{j+r-\rho} - \integ{j-s+\sigma}
+ (1-\rho)\sigma  \comma \nonumber \\
{\tilde{\alpha}}'_j & = & \integ{j+r-\rho} - 
\integ{j-s-\rho \sigma}
+ \rho (1-\sigma) \comma \nonumber \\
{\tilde{\beta}}_j & = & - \integ{j+r-(1-\rho)(1-\sigma)} +
\integ{j-s - 1+ \sigma  } 
+ \rho (1-\sigma) \comma \nonumber\\
{\tilde{\beta}}'_j & = &  \integ{j+r-s -(1-\rho)\sigma  } - 
\integ{j }  + (1-\rho)\sigma
 \comma  \\
{\tilde{\gamma}}_j & = &  \integ{j+r-\rho } + \integ{j-s+ \sigma }
 - \integ{j } - \integ{j+r-s } -  (1-\rho) \sigma + \rho (1-\sigma)
(-1+ 2\, \delta_{j,s-r}) 
 \comma \nonumber \\
{\tilde{\rho}}_{ij} & = & \delta_{i,0} \delta_{j,s-\sigma} + 
\delta_{j,0} \delta_{i,r-\rho} - \delta_{i,0}\delta_{j,0} \period \nonumber  \ea
From (\ref{defmunu}), incrementing $r$ ($s$) by $N$ increments $\mu$ ($\nu$) by 1,
so the right-hand sides of (\ref{reslts}) are each  periodic in  $r$, $s$ and $j$, 
of period $N$, as they should be.

\noindent 
Define
\be \phi_{\rho \sigma } \eq \rho+\sigma - 2 \rho \sigma \comma \ee
so that $\phi_{\rho \sigma }  = 0$ if  $ \rho = \sigma $, and 
$\phi_{\rho \sigma }  = 1$ if  $ \rho \neq \sigma $. 
Our numerical experiments also suggest  that,
\ba \label{sumsa} 
\sum_j \alpha_j & = & \sum_j {{\beta}_j}' \eq (N-1) (1-\rho) \sigma + 
\, \sum_j (n_j - m_j) \comma \nonumber \\
\sum_j {{\alpha}_j}' & = & \sum_j {\beta}_j \eq  (N-1) \rho(1- \sigma) + 
\, \sum_j (m_j - n_j) \comma  \\
\sum_j {{\gamma}_j} & = &  - (N \minus 1) \,  \phi_{\rho \sigma }  \comma \nonumber  \ea
the sums being from $j=0$ to $j=N-1$.  It follows that
\ba \label{sumsb}
\sum_j (\alpha_j + {\alpha_j}' + \gamma_j ) & = &  \sum_j (\beta_j + 
{\beta_j}' + \gamma_j ) \eq 0 \comma  \\
\sum_j (\alpha_j + {\beta_j} + \gamma_j ) & = &  \sum_j ({\alpha_j}' + 
{\beta_j}' + \gamma_j ) \eq 0 \period \nonumber \ea

\noindent We also find that
\be \label{sumsc}
\sum_j j (\alpha_j + {\alpha_j}' + \beta_j + 
{\beta_j}' +  \gamma_j ) \eq 0 \sep {\rm mod} \; N \ee
if $N$ is odd or if $\rho = \sigma$; if $N$ is even and $\rho \neq \sigma$, then 
the rhs of (\ref{sumsc}) 
equals $N/2$, modulo $N$.

These relations (\ref{sumsa}) - (\ref{sumsc}) ensure that no additional external 
constant factors occur in (\ref{Ars}). 
For instance, multiplying $a_p$ by $\omega^k$, will introduce a factor
$\omega^{k L}$, where $L = \sum_j (\alpha_j + {\alpha_j}' +\gamma_j )$. 
From (\ref{sumsb}), this 
factor is  unity. Similarly for $b_p, a_q, b_q$. Also,
interchanging all $p$ variables with the corresponding $q$ variables will introduce a factor
$(-1)^I \, \omega^J$, where
\bd
I = \sum_j ( \alpha_j +{\alpha_j}' + \beta_j +  {\beta_j}' + \gamma_j ) \sep J = 
\sum_j j ( \alpha_j +{\alpha_j}' + \beta_j +  {\beta_j}' + \gamma_j ) \period \ed
From (\ref{sumsa}) and (\ref{sumsc}),  this factor is also unity.

If we ignore the requirement (\ref{defmunu}),\footnote{Presumably there is some variant of 
this symmetry that preserves the equations (\ref{reslts}) completely.},
the relations (\ref{reslts1}) are unchanged by the substitutions
$\rho,\sigma,r,s,\mu, \nu, m_j,$ $ n_j, \alpha_j, \alpha'_j,
 \beta_j, \beta'_j,\gamma_j,\rho_{ij} \rightarrow 1-\sigma, 1-\rho,1-s,1-r,-\nu,
 -\mu, - n_{1-j}, -m_{1-j} ,$ $ \alpha_j, \beta_{j+1},
 \alpha'_{j-1}, \beta'_j,\gamma_j,\rho_{-j,-i} $.

\subsubsection*{``Dimension'' of the Riemann surface }

We see that the $2 N$ integers $m_0, \ldots , m_{N-1}, n_0, \ldots , n_{N-1}$ are 
sufficient to specify the function $T_{pq}$ on any sheet within its Riemann surface,
so any sheet can be associated with a point in a $2N$-dimensional space. Further,
incrementing each of  $m_0, \ldots , n_{N-1}$ (and therefore also $\mu ,\nu$) by unity 
(or any integer) leaves (\ref{reslts}) unchanged, so the space can be reduced 
to one of dimension $2N - 1$. Ths appears to be a partial analogue for the chiral Potts model
of the rapidity ``difference property'' that plays such an important role in the simpler 
models.

For $N > 2$ this appears to be the best one can do - each sheet of the Riemann surface 
corresponds to a point in a $(2N-1)$-dimensional space. Note however that  sheets 
are neighbours if their $m$-integers satisfy (\ref{mnbrs}),  or if 
their $n$-integers satisfy (\ref{mnbrs}). Hence neighbouring sheets do not necessarily
correspond to neighbouring points in $m_0, \ldots , n_{N-1}$-space.

For $N=2$ we have the additional 
relations $m_{1-\rho } = n_{1-\sigma} = 0$ for all sheets, so $m_0, m_1, n_0$, $n_1$ enter
(\ref{reslts}) only via $m_{\rho} - n_{\sigma}$ and the space is  
one-dimensional, as we observed in \cite{paper1}.

For $N>2$ we note that the exponents $\alpha_j, \ldots , \beta'_j, \rho_{ij}$ 
are linear in the integers $m_0, \ldots , n_{N-1}$, but the $\gamma_j$ are linear 
only separately in $m_0, \ldots , m_{N-1}$ and $n_0, \ldots , n_{N-1}$. They are bilinear
in the full set of $2N$ integers, due to the occurrence of the $g_j$, as defined by 
(\ref{defgint}).

The values of $p$ = $\{ x_p, y_p, \rho, r, m_0, \ldots, m_{N-1} \}$ completely specify
the point $p$, not only on the algebraic curve (\ref{curve}), but also on
the Riemann surface of $T_{pq}$. Similarly, $q$ = $\{ x_q, y_q, \sigma, s,$ $ n_0, \ldots,
 n_{N-1} \}$ completely specifies $q$.
We shall refer to the corresponding Riemann sheet 
as the ``sheet $(m,n) \, $''.

\section{Zeros and poles of $T_{pq}$ }

For $a_p, b_p, a_q, b_q$ in the central domain $\cal D$, the functions 
$T(\omega^i a_p, b_p | \omega^j a_q, b_q)$ on the rhs of 
(\ref{Ars})  are non-zero and analytic. Hence the rhs of (\ref{Ars}) is meromorphic
in $\cal D$, with zeros or poles only when
\be a_p = \omega^j a_q \; \; \; {\rm and} \; \; \; b_p = b_q \comma \ee
which implies $ a_p b_p= \omega^j a_q b_q$. Thus its zero at this point is of order
\be
 \alpha_j + {\beta}'_0 + \gamma_j \period \ee
(Equivalently, its pole is of order $-\alpha_j - {\beta}'_0 - \gamma_j$.)

From (\ref{curve}), the relation $x_p^N = x_q^N$ implies $y_p^N = y_q^N$ and vice-versa, 
and either implies
$t_p^N = t_q^N$. Similarly, $x_p^N = y_q^N$ implies $y_p^N = x_q^N$ and again 
$t_p^N = t_q^N$. From (\ref{defA}) it follows that the function $T(x_p,y_p|x_q,y_q)$ is meromorphic 
throughout  
its Riemann surface, with zeros and poles only when $t_p^N = t_q^N$. 
On the sheet $(\rho,\sigma,r,s,m,n)$ 
the zero at $t_p = \omega^k t_q$ ($k = 0, \ldots , N \! - \! 1$) is of order
\bd z_k \eq (-1)^{\rho+\sigma} \,  (\alpha_{k+s-r} + {\beta}'_0 + \gamma_{k+s-r} )\period  \ed
From (\ref{reslts}) and (\ref{reslts1}) it follows that
\be \label{zeros}
(-1)^{\rho+\sigma}  z_k \eq 2 g_{k+1} - 2 g_{k-1} - 2 m_{k+\sigma} + 2 n_{\rho-k} + 2 
\rho (1-\sigma) (\delta_{k,0} - 1) + w_k \comma \ee
where
\be w_k \eq \integ{k+s-r-\rho(1-\sigma)} - \integ{k} + 
\integ{r-s-(1-\rho) \sigma } + \phi_{\rho \sigma}  \period \ee

The contribution $w_k$ has a simple explanation. Using the notation of 
\cite{Seoul90} and \cite{Baxter93}, let us postulate the existence of  
functions
$\Theta_{i,j}$, 
 $\overline{\Theta}_{i,j}$ of $p$ and $q$ such that $\Theta_{i,j}$ has simple zeros on 
the Riemann surface when
\be \label{oThetadefn}
x_q = \omega^i y_p \; \; {\rm and } \; \; y_q = \omega^j x_p \comma \ee
and $\overline{\Theta}_{i,j}$ has simple zeros when
\be \label{Thetadefn}
x_q = \omega^i x_p \; \; {\rm and } \; \; y_q = \omega^j y_p \period \ee
Any zero or pole of $T_{pq}$ must occur at one of these points on some 
particular Riemann sheet.

Define $G_{pq}$ by
\be \label{defG}
 G_{pq}  \eq G(x_q,y_q) \eq  \prod_{i=1}^{N-1} \prod_{j= 1}^{N-i}  
\left[ \Theta_{N-i,N+1-j} \, \overline{\Theta}_{i,j}  \right] ^{-1} \comma \ee
Then $G_{pq}$ is the function $1/\xi \overline{\xi}$ of  \cite{Seoul90}. We remark 
therein that its poles contain just all the poles of the Boltzmann weight functions
$W_{pq} (n), \overline{W}_{pq}(n)$. Hence for a finite square lattice of 
$\cal N$ sites with partition function $Z$, it is true that 
$Z/G_{pq}^{\cal N} $ has no poles on the Riemann surface.

It is therefore natural to define a normalized function $\tilde{T}_{pq}$ by 
\be \label{defTT}
 T_{pq} \eq    G_{pq}  \tilde{T}_{pq} \comma \ee
since  this $\tilde{T}_{pq}$ is the partition function per site in 
this pole-free normalization.

On the sheet $(\rho,\sigma,r,s,m,n)$ 
we find from (\ref{defG}) that the pole of $G_{pq}$ at $t_p = \omega^k t_q$ ($k = 0, \ldots , N \! - \! 1$) 
is of order $- (-1)^{\rho+\sigma } \, w_k$ 
(so numerically this is either 0 or 1).
  Hence the term $w_k$ in (\ref{zeros}) is 
the contribution of $G_{pq}$. The preceding terms (all even integers) are the contribution of
$\tilde{T}_{pq}$.

Let $\theta_k$ be the function with simple zeros on the Riemann surface when
$t_p = \omega^k t_q$ and $x_p^N = y_q^N$, $y_p^N = x_q^N$ . Then 
\be \theta_k \sim \prod_{i=0}^{N-1} \Theta_{i,-k-i} \comma \ee
where by $f \sim g$ we mean herein that $f$ and $g$ have the same 
zeros and poles (when $t_p^N = t_q^N$) on the Riemann surface.

Similarly, 
\be \label{lastth}
\overline{\theta}_k \sim \prod_{i=0}^{N-1} \overline{\Theta}_{i,-k-i} \comma \ee
where $\overline{\theta}_k$ has simple zeros when  $t_p = \omega^k t_q$ and $x_p^N = x_q^N$,
$y_p^N = y_q^N$. Then 
  formally we can write (\ref{zeros}) as
\be \label{TTpqz}
\tilde{T}_{pq} \sim \prod \left( \overline{\theta}_k / \theta_k   \right)^{2 \epsilon_{pq}(k)}
\comma \ee
where the product is over all Riemann sheets $(\rho,\sigma,r,s,m,n)$ and all 
values $0,\ldots ,N \! - \! 1$ of $k$, and the integer $\epsilon_{pq}(k)$ is 
\be \label{defVpq}
\epsilon_{pq}(k) \eq  g_{k+1} -  g_{k-1} -  m_{k+\sigma} +  n_{\rho-k} +  
\rho (1-\sigma) (\delta_{k,0} - 1) \period \ee
If $\rho = \sigma$ then only 
$\overline{\theta}_k$ can have zeros, and if $\rho \neq \sigma$ only 
$\theta_k$.

We note that there are various formal identities between our $\Theta$ and $\theta$ functions, notably
\bd \prod_j \overline{\Theta}_{ij} \sim x_q - \omega^i y_p  \sep
\prod_i \overline{\Theta}_{ij} \sim y_q - \omega^j x_p  \sep 
\prod_j \Theta_{ij} \sim x_q - \omega^i x_p  
  \ed
\be \label{threlns}
\prod_i \Theta_{ij} \sim y_q - \omega^j y_p\sep  \prod_k \theta_k  \overline{\theta}_k
\sim t_p - \omega^k t_q \comma \ee
all products being over the integers $0, \ldots , N \! - \! 1$.

It may be possible to give explicit representations of our postulated functions 
$\Theta_{ij}, \overline{\Theta}_{ij},
G_{pq}$, $\theta_k, \overline{\theta}_k, \tilde{T}_{pq}$ in terms of hyperelliptic functions
 \cite{Kyoto,rjb93,rjb98}, but we shall not do so here. Using them does greatly 
simplify the relations and certainly provides a way of keeping track of the poles and zeros at 
$t_q^N = t_p^N$ on the Riemann surface. These are the only poles and zeros, apart possibly from 
ones occuring when $p$ has some particular value independent of $q$, or $q$ has 
some value independent of $p$. This suggests that herein the relation
\bd f_{pq} \sim g_{pq} \ed
is equivalent to the explicit identity 
\bd
\frac{f_{pq} f_{p'q'}}{f_{pq'} f_{p'q}} = \frac{g_{pq} g_{p'q'}}{g_{pq'} g_{p'q}} \comma \ed
for all rapidities $p,q,p',q'$. This implies that 
there exist single-rapidity functions $u_p, v_q$ such that 
\bd f_{pq} = u_p g_{pq} v_q \period \ed
In all the cases where we have been able to test this hypothesis, e.g. by using 
(\ref{threlns}), we have found it to be true.

\section{Automorphisms}

Three basic automorphisms that take a point on on the curve (\ref{curve}) 
to another such point
are $R, S, U$, where
\be \label{RSU}
R : \; \; x_{Rp} = y_p \sep y_{Rp} = \omega x_p \; \; \; ; \; \; \;  S: \; \; x_{Sp} = y_p^{-1} 
\sep y_{Sp} = x_p^{-1} \ \; ;  \ee
\bd U: \; \; x_{Up} = \omega x_p \sep y_{Up} =  y_p \period \ed 
If $p$ lies in the initial domain  ${\cal D}_0$ (so $y_p \simeq 1$), then we can take $Up$ 
to also lie in   ${\cal D}_0$; and $Rp$, $Sp$ 
 to lie in the adjacent domain ${\cal D}'_0$.\footnote{$R$ and $S$ are the same as in 
\cite{BPAY88}, while $R, U$ occur in \cite{BBP90}.}

We can determine what happens when $p$ is not in the initial domain ${\cal D}_0$
by analytic continuation. If $p$ lies on a sheet that is a $k$th neighbour of 
${\cal D}_0$, a  route to it being  the sequence (\ref{seq}) of sheets of types 
$\{ 0,r_1,r_2, \ldots, r_k \}$, then $Up$ is also on a $k$th neighbouring sheet 
(so has the same parity $\rho$), but from (\ref{RSU}) the sequence to $Up$ is 
$\{ 0,r_1+1,r_2,r_3+1, \ldots, r_k + \rho \}$.\footnote{Note that in general $Up$ 
is {\em not} obtained from
$p$ by staying on the same sheet and simply replacing $x_p$ by $\omega x_p$.}

As above, writing $p = \{ x_p, y_p, \rho, r, m_j \}$ for the full set
of parameters defining the rapidity $p$ (with $j = 0, \ldots , N \minus 1$), it follows that
\be \label{defU}
 Up \eq  \{ \omega x_p, y_p, \rho, r+ \rho , m_{ j-1} + (-1)^\rho (m_{ N-1} -
 m_{  0}) - \rho \,  \delta_{j,1} \, 
 \} \period \ee
Iterating , we obtain for all integers $i$
\be U^i p \eq  \{ \omega^i x_p, y_p, \rho, r+ i\rho , m_{ j-i} + (-1)^\rho (m_{ N-i} -
 m_{  0} ) - \rho \,  F_{j,i} \,  \} \period \ee
The $r$ values should always be taken modulo $N$, so that $0 \leq r < N$. Then we see, 
using (\ref{periodic}), that  $U^N p = p$.

Similarly, $Rp$ and $Sp$  lie on  $(k+1)$-th neighbouring sheets of ${\cal D}_0$, 
at the termini of
the routes $\{ 0,0,r_1+1,r_2,r_3 +1, \ldots, r_k + \rho \}$, 
 $\{ 0,0,-r_1,-r_2,-r_3, \ldots,- r_k  \}$, respectively, and
\be \label{autoR}
Rp \eq  \{ y_p,\omega x_p, 1 - \rho, r+ \rho , m_{ j-1}  + \rho (1 \minus 
  \delta_{j,1} )
 \, \} \comma \ee
\be \label{autoS}
Sp \eq  \{ 1/y_p,1/x_p, 1 - \rho, - r , - m_{ N+1-j}  \, \} \period \ee

Two combinations of automorphisms that we shall need are $V = R U^{-1}$ and $U^i V$:
\be
V p \eq R U^{-1} p \eq \{  y_p, x_p, 1-\rho, r, 
m_{j}+(-1)^\rho (m_1 - m_{0})  \, \} \ee
and
\be
U^i V p \eq \{ \omega^i y_p, x_p, 1-\rho, r+i(1-\rho), 
m_{j-i}+(-1)^\rho (m_1 - m_{N-i}) - (1-\rho) F_{j,i} \, \} \ee
for any integer $i$.

The effect of these automorphisms on the second rapidity $q$ can be obtained at once
by replacing $p, \rho, r, m_0, \ldots , m_{N-1}$ by $q, \sigma, s, n_0, \ldots , n_{N-1}$.

\section{Relations for $\tilde{T}_{pq}$ and its exponents}

Going to the pole-free normalization (\ref{defTT}), the  relations
(\ref{Trcontp}), (\ref{Trcontq}), (\ref{2ndratio}) simplify to
\be \label{Trcontp1}
\tilde{T}_r(\omega^r y_p, x_p|x_q,y_q )  \sim   \frac{\tilde{T}(\omega^r x_p,y_p|x_q,y_q ) \; 
 \overline{\theta}_{1}^2
\overline{\theta}_{2}^2 \cdots \overline{\theta}_{N-r-1}^2 \; \theta_{N-r}^2  \theta_{s+1}^2 
\cdots   \theta_{N-1}^2  }
{\tilde{T}(\omega^{-1} x_p,y_p|x_q,y_q ) \, \tilde{T}( x_p,y_p|x_q,y_q )} 
 \ee
 \be \label{Trcontq1}
\tilde{T}_s(x_p,y_p|\omega^s y_q, x_q)  \sim   \frac{\tilde{T}(x_p,y_p|\omega^s x_q,y_q) \;
 \overline{\theta}_{0}^2
\overline{\theta}_{1}^2 \cdots \overline{\theta}_{s-1}^2 \; \theta_s^2  \theta_{s+1}^2 \cdots 
\theta_{N-2}^2  }
{\tilde{T}(x_p,y_p|\omega^{-1} x_q,y_q) \tilde{T}(x_p,y_p| x_q,y_q)} \ee 
\be \label{2ndratio1}
\frac{\tilde{T}(x_p, y_p| x_q, y_q) \, \tilde{T}(\omega x_p, y_p| \omega x_q, y_q)}{
\tilde{T}(x_p, y_p|\omega  x_q, y_q) \, \tilde{T}(\omega x_p, y_p|  x_q, y_q)} \sim 
\left( \theta_{N-1} / \theta_0 \right) ^2 \period \ee

Again, in these relations as written, $p, q$ both lie in the central domain ${\cal D}_0$.
However, we can now analytically continue to any Riemann sheet and use the above automorphisms
to obtain
\bd 
\tilde{T}_{U^rVp,q} \sim   \tilde{T}_{U^rp,q} \; 
 \overline{\theta}_{1}^2
\overline{\theta}_{2}^2 \cdots \overline{\theta}_{N-r-1}^2 \; \theta_{N-r}^2  
\theta_{N-r+1}^2  \cdots   \theta_{N-1}^2  /
\left( \tilde{T}_{U^{-1}p,q} \, \tilde{T}_{pq} \right) \comma  \ed
 \be \label{Trcontq2}
\tilde{T}_{p, U^s Vq}  \sim  \tilde{T}_{p,U^s q} \;
 \overline{\theta}_{0}^2
\overline{\theta}_{1}^2 \cdots \overline{\theta}_{s-1}^2 \; \theta_s^2  \theta_{s+1}^2 \cdots 
\theta_{N-2}^2  /\left(
\tilde{T}_{p,U^{-1} q} \tilde{T}_{pq} \right) \comma \ee 
\bd 
\tilde{T}_{pq} \, \tilde{T}_{Up,Uq} /(\tilde{T}_{p,Uq} \, \tilde{T}_{Up,q} ) \sim 
\left( \theta_{N-1} / \theta_0 \right) ^2  \ed
for all $p,q$. Note that in these relations $r,s$ must both have values in the set 
$\{ 0, 1, \ldots , 
N \minus 1 \}$.

Substituting the form (\ref{TTpqz}) of $\tilde{T}_{pq}$, we obtain the exponent relations
\bd
-\epsilon_{U^r Vp,q}(k+r) = \epsilon_{U^r p,q}(k+r) - \epsilon_{U^{-1}p,q}(k-1) - \epsilon_{pq}(k) + 
\integ{k+(N \minus 1) (1-\phi_{\rho \sigma ) }} - \integ{k+r} +\integ{r} \comma \ed
\be \label{exprelns}
-\epsilon_{p,U^s Vq}(k-s) = \epsilon_{p,U^s q}(k-s) - \epsilon_{p,U^{-1} q}(k+1) - \epsilon_{pq}(k) + 
\integ{k+ (N \minus 1) \phi_{\rho \sigma }} - \integ{k-s} -\integ{s}\comma \ee
\bd
\epsilon_{pq}(k) + \epsilon_{Up,Uq}(k) - \epsilon_{Up,q}(k+1) - \epsilon_{p,Uq}(k-1) \eq \phi_{\rho \sigma}
(\delta_{k,0} - \delta_{k,N-1}) \period \ed
We have used Mathematica to verify for $N = 2, \ldots, 12$ that  these equations are 
indeed satisfied by (\ref{defVpq}). They are explicitly periodic in $r,s$, of period $N$, 
so are true for all integers $r,s$.

For future reference, we note that the half-exponents for $\tilde{T}_{p,U^{\alpha} q}$  are
\be  \epsilon_{p,U^{\alpha} q} (k-{\alpha}) \eq \psi(\rho, \sigma, k,{\alpha}|m,n )  \comma \ee
where
\bd
\psi(\rho, \sigma, k,{\alpha}|m,n )  \eq  g_{k+1}-g_{k-1} +n_{\rho-k} - 
\sigma ( m_{k} + m_{k+1} )
 +
(-1)^{\sigma} (n_{N-{\alpha}} - n_0  - m_{k-{\alpha}} ) -  \ed \be \label{psidef} 
\sigma F_{\rho-k+{\alpha},{\alpha}} + \rho(1-\sigma) (\delta_{k,{\alpha}} - 1)
\period \ee

Also, those for and 
$\tilde{T}_{U^{\alpha} p, q}$ are
\bd  \epsilon_{U^{\alpha} p,q} (k+{\alpha}) \eq g_{k+1}-g_{k-1} - m_{k+\sigma} +
\rho ( n_{-k} + n_{1-k} )
 + 
(-1)^{\rho} (m_0 - m_{N-{\alpha}}   + n_{-k-{\alpha}} ) + \ed \be \label{st2}
\rho F_{k+{\alpha}+\sigma,{\alpha}} + \rho(1-\sigma) (\delta_{k,-{\alpha}} - 1)
\period \ee

\section{The function $\tau_2(p,q)$}
In the derivation of $T_{pq}$ given in (\cite{Seoul90}) and (\cite{Baxter90}), 
an important 
role is played by the auxiliary function $\tau_2(p,q)$. Noting that the $T(x_q,y_q)$ in 
(\cite{paper1})
is $T_{pq}$, while $T(\omega x_q,y_q)$ is $T_{p,Uq}$, it follows that 
equation (25) of (\cite{paper1})
can be written, for all Riemann sheets,  as 
\be \label{tau2eqn}
\tau_2 (p,q) \eq \left[ \frac{
 (y_p - \omega x_q ) (t_{p} - t_q)}{y_p^2 \, (x_{p} - x_q) } \right]^L 
 \; \frac{ T_{pq} }{T_{p,Uq} }  \ee
writing $\tau_2(t_q)$ in (\cite{paper1}) as $\tau_2(p,q)$. Also, 
in (53) of \cite{paper1} we obtain the result
\be \label{tau2result}
\log \tau_2 (p,q) \eq \frac{1}{2\pi} \, \int_{0}^{2 \pi} \left( \frac{1 + \lambda_p e^{i \theta}}
{1 - \lambda_p e^{i \theta}} \right) \, \log \! \left[ \frac{\Delta(\theta) - \omega t_q }{y_p^2} 
\right] \, d\theta \ee
for $p,q$ both in the central domain ${\cal D}_0$, where $|\lambda_p|, |\lambda_q| < 1$. Here
\be  \label{defDel}
\Delta (\theta) \eq \left( \frac{1 -2 k' \cos \theta + {k'}^2}{k^2} \right)^{1/N} \ee

Setting $j=N$ in (14) of (\cite{paper1}) and using (21) and (23) therein,we find
\be
T_{pq} T_{p,Vq} \eq 
\frac{N  \, (y_p - x_q) \, (y_p - y_q) \, (t_p^N-t_q^N) }
{y_p^2 \, (x_p^N - x_q^N) \, (y_p^N-y_q^N) \, \tau_2 (p,U^{-1}q )}   \ee
for all Riemann sheets. The LHS is the free energy (per double site) of a model with vertical rapidities $p$ and 
alternating horizontal rapidities $q$, $Vq$. This is the general  ``superintegrable''
model discussed in \cite{Baxter89}, so we see that  to within simple known scalar factors, 
$1/\tau_2(p,U^{-1} q)$ is the free energy of this model. Indeed $\tau_2(p,U^{-1} q)$ itself
is the free energy (apart possibly from simple scalar factors) of the ``inverse'' model introduced in 
\cite{Baxter89}, while $\tau_2(p,q)$ is the free energy of the model defined in (3.44) - (3.48) of 
\cite{BBP90} (with $j=2$ and $k=0$).

We can take (\ref{tau2eqn}) as the definition of $\tau_2(p,q)$. It follows at once
that
\be \label{tau2reln}
\tau_2(p,q) \tau_2(p,Uq)  \cdots \tau_2(p,U^{N-1} q) \eq  \frac{ (y_p^N- x_q^N ) (t_{p}^N- t_q^N) }
{y_p^{2N}  (x_{p}^N- x_q^N) } \ee
  
Also, considered as a function of $t_q$,  $\tau_2(p,q)$ only has a single branch cut, 
from
$\omega^{-1} \eta$ to $\omega^{-1}/\eta$. Across the other $N-1$ potential branch cuts it is in fact an
analytic function of  $t_q$. Together with (\ref{tau2reln}), this implies that
\be \label{tau2contn}
\tau_2(p,U^i V U^{-i} q) \eq v_{pq}^{\delta_{i,N-1}} \, \tau_2 (p,q) \comma \ee
where
\be v_{pq} \eq \frac{(x_p^N-x_q^N) \, (y_p^N-y_q^N)}
{(x_p^N-y_q^N) \, (y_p^N-x_q^N)} \period \ee
(Going from $q$ to $U^i V U^{-i} q$ takes one from ${\cal D}_0$ to the neighbouring 
domain ${\cal D}'_i$, while leaving $t_q$ unchanged.)

One can also verify directly from (\ref{tau2reln}) that
\be \label{tauprod} \tau_2(p,q)  \,  \tau_2(U^i V U^{-i} p,q)\eq 
(\omega^{-i} t_p  - \omega t_q)^2/y_p^4 \period \ee

\subsubsection*{Exponent relations}
The exponents of $\tau_2(p,q)$ are simpler than those of the free energy function
$T_{pq}$, 
from  (\ref{tau2eqn}) and (\ref{TTpqz}) ,
\be  \tau_2(p,q)  \sim  \theta_0^2 \, \tilde{T}_{pq} /\tilde{T}_{p,Uq} 
 \sim  \prod \left( \overline{\theta}_k /
 \theta_k \right) ^{2 \tilde{\epsilon}_{pq}(k)} \comma    \ee
where
\ba \label{epsvals}
\tilde{\epsilon}_{pq}(k) & = &  \epsilon_{pq}(k) - \epsilon_{p,Uq}(k-1) - 
\phi_{\rho \sigma} \delta_{k,0} \nonumber \\
& = & (-1)^\sigma \left( m_{k-1} - m_k +n_0 - n_{N-1} - \rho \delta_{k,1} \right) \period
 \ea
We see that the bilnear terms $g_k$ do not occur in $\tilde{\epsilon}_{pq}(k)$, leaving only 
terms that are fully linear in the $m_j$ and $n_j$. 

 The equations (\ref{tau2reln}), (\ref{tau2contn}), (\ref{tauprod} ) imply that 
the exponents  $\tilde{\epsilon}_{pq}(k)$ satisfy
\bd \tilde{\epsilon}_{pq}(k) + \tilde{\epsilon}_{p,Uq}(k-1) + \cdots 
+ \tilde{\epsilon}_{p,U^{N-1} q}(k-N+1)  =  - \, \phi_{\rho \sigma} \comma \ed
\be  -\tilde{\epsilon}_{p, U^i V U^{-i} q} (k)  \eq  \tilde{\epsilon}_{p q}(k) + 
\delta_{i,N-1} \comma \ee
\bd \tilde{\epsilon}_{p q}(k) - \tilde{\epsilon}_{U^i V U^{-i} p, q} (k) \eq  
(1-2 \, \phi_{\rho \sigma}) 
\, \delta_{k,i+1}    \ed
for all $\rho, \sigma, k, m_0, \ldots , m_{N-1}, n_0, \ldots ,n_{N-1}$. Indeed we find, 
using (\ref{epsvals}) and (\ref{defU}), that this is so.


The half-exponents for $\tilde{\epsilon}_{p, U^{\alpha}  q} $ are

\be \label{st3} \tilde{\epsilon}_{p,U^{\alpha} q} (k-{\alpha}) \eq (-1)^{\sigma} ( m_{k-{\alpha}-1} -
 m_{k - {\alpha}} +
n_{-{\alpha}} - n_{-{\alpha}-1}
-\rho \delta_{k,{\alpha}+1} + \sigma \delta_{{\alpha},N-1} ) \period \ee

\section{``Sufficiency'' of the rotation and inversion relations}

As is consistent with series expansions, the  function $T_{pq}$ is
non-zero and analytic in the central (physical) domain ${\cal D}_0$.
Given this, the recursion relations (\ref{Trcontp}) - (\ref{2ndratio}) are certainly 
sufficient to determine 
the Riemann surface on which  $T_{pq}$ lives, that $T_{pq}$ is meromorphic on 
this surface, and to give  the orders (exponents) $\epsilon_{pq}(k)$ of all 
its zeros and poles on every sheet.

This goes a long way towards defining $T_{pq}$. To complete the description
one needs to establish that the ratio of two such functions with the same zeros 
and poles has some periodicity property from sheet to sheet so that it is bounded 
over the whole surface. It is certainly entire, so by Liouville's theorem it would 
then have to 
be a constant. Such constants can usually be fixed from special cases.

We shall not discuss this problem of completing the description further herein, but 
will suppose that it can be done. Here our concern is to see if the weaker 
rotation and inversion relations can be used to determine the exponents
$\epsilon_{pq}(k)$. 

More specifically, the  rotation and inversion
relations, together with the analyticity properties in 
the central domain, are known to be sufficient to determine the free energy
for the two-dimensional lattice models with the ``rapidity difference property''.
\footnote{This property implies that such a model can be parametrized in terms of 
single-argument Jacobi elliptic functions, which fixes the Riemann surface for the 
free energy.}

The chiral Potts model does not possess the rapidity difference property, and there is no
parametrization in terms of single-argument Jacobi elliptic functions. Nevertheless,
we can ask whether its rotation and inversion
relations, together with some simple and plausible ansatz,
are sufficient to determine the free energy exponents $\epsilon_{pq}(k)$. This is 
the question we address in the remainder of this paper.

\subsubsection*{The relations}
From equations 39, 40 and 10 of \cite{paper1}, the rotation and inversion relations
are
\be \label{rotn} T_{q,Rp} \eq T_{pq} \sep
 T_{qp} T_{pq} \eq r_{pq} \comma \ee
where \be r_{pq} \eq \frac{N \, (x_p-x_q) \, (y_p - y_q) (t_p^N - t_q^N) }
{(x_p^N -x_q^N ) \, (y_p^N  - y_q^N ) (t_p - t_q) } \period \ee

From these we can deduce that
\be T_{Rp,q} T_{pq} = r_{Rp,q} \sep T_{p,Rq} T_{pq} = r_{pq} \period \ee
These two equations can be obtained from (\ref{Trcontp}), (\ref{Trcontq}) by setting 
$r = s = 0 $ and replacing either $x_p$ or $ x_q$ by $\omega x_p$ or $
 \omega x_q$.
They can also be obtained by setting $r = s = N-1$ and replacing either $p$ or $q$ by 
$Rp$ or $Rq$. There is therefore an overlap between (\ref{rotn}) and the recursion relations
 (\ref{Trcontp}), (\ref{Trcontq}), but the latter can not be deduced from the former.

Using (\ref{defG}) and (\ref{defTT}), (\ref{rotn}) become
\be \label{rotna} \tilde{T}_{q,Rp} \sim \tilde{T}_{pq} \sep
 \tilde{T}_{qp} \tilde{T}_{pq} \sim  
(\theta_1 \theta_2 \cdots \theta_{N-1} )^2 \period \ee

On a given Riemann sheet of $p$-parity and type $\rho, r$,  and 
$q$-parity and type $\sigma, s$, specified by the integers
$m_0, \ldots , m_{N-1}$, $n_0, \ldots , n_{N-1}$, let the order of the zero 
when $t_p = \omega^k t_q$ be
$2\,  (-1)^{\rho+\sigma} e(\rho,\sigma, k,r,s|m, n )$, where $m = \{ m_0 ,\ldots, m_{N-1} \}$
and  $n = \{ n_0 ,\ldots, n_{N-1} \}$.
Then, analogously to (\ref{TTpqz}), we can write 
\be \label{prodform}
\tilde{T}_{pq} \sim \prod \left( \overline{\theta}_k/\theta_k \right)^ {2 
 e(\rho,\sigma, k,r,s|m, n )} \comma \ee
the product being over all zeros (and poles), and all Riemann sheeets.

Are the rotation and inversion relations (\ref{rotna}) sufficient to fix 
$ e(\rho,\sigma, k,r,s|m, n )$ as the $\epsilon_{pq}(k)$ given by (\ref{defVpq})?
First we must note that there are severe self-consistency restrictions on the exponents
$ e(\rho,\sigma, k,r,s|m, n )$, for any meromorphic function on the Riemann surface.

\subsubsection*{Consistency}

Consider some particular zero, at $t_p = \omega^k t_q$,  on some particular sheet of types 
$r, s$. For definiteness, 
take the parities $\rho, \sigma$ to be zero. Then (since $y_p \simeq \omega^r, y_q \simeq \omega^s$) the zero is at 
$x_p = \omega^{k+s-r} x_q$, $y_p = \omega^{r-s} y_q$.

Now move $p$ and $q$ to  adjacent sheets of types $r', s'$, respectively, so their
parities both become one. Now consider the zero at  $t_p = \omega^{k'} t_q$. This must be
when $x_p = \omega^{r'-s'} x_q$ and $y_p = \omega^{k'+s'-r'} y_q$. If $k = k' = r+r'-s-s'$,
this is the same zero as the one on the previous sheet: we have simply followed it from 
one sheet to the next. Its exponent (an integer) must be the same, so
\be \label{cons}
e(\rho,\sigma, k,r,s|m, n ) \eq e(1-\rho,1-\sigma, k,r',s'|m', n' )\comma \ee
provided $k = r+r'-s-s'$ and ${m'}, {n'}$ are given by (\ref{mnbrs}), (\ref{nnbrs}).
We have only established this condition for $\rho = \sigma = 0$, but the same result is 
obtained for all $\rho, \sigma$.

This is a very strong condition on the exponents $e(\rho,\sigma, k,r,s|m, n ) $. It 
must be true for all values of $\rho, \sigma, r,s, r',s'$, and all values of the integers
$m_j, n_j$. Furthermore , it must be true for any meromorphic function on the Riemann surface.
Hence it is true for the exponents $\tilde{\epsilon}_{pq}(k)$ of $\tau_2 (p,q)$
as well as those of $T_{pq}$ and $\tilde{T}_{pq}$.

\subsubsection*{Rotation and inversion}

From (\ref{oThetadefn}) - (\ref{lastth}), exhibiting the dependence of $\theta_k$,
$\overline{\theta}_k $ on $p, q$:
\be \left[\theta_k \right]_{qp} \eq  \left[\theta_{-k} \right]_{pq} \sep
\left[\overline{\theta}_k \right]_{qp} \eq  \left[\overline{\theta}_{-k} \right]_{pq} \comma \ee
 \be \left[\theta_k \right]_{q,Rp} \eq  \left[\overline{\theta}_{-k-1} \right]_{pq} \sep
\left[\overline{\theta}_k \right]_{q,Rp} \eq  \left[\theta_{-k-1} \right]_{pq} \period 
\ee

\noindent Together with (\ref{prodform}), (\ref{rotna}) and (\ref{autoR}), these imply 
the relations
\be \label{rotnb}
e(\rho,\sigma, k,r,s|m, n ) \eq - e(\sigma,1 - \rho,  -k-1,s, r+\rho |n, Rm
) \comma \ee
\be \label{invnb}
e(\rho,\sigma, k,r,s|m, n ) +  e(\sigma, \rho, -k,s, r |n,
 m ) \eq \phi_{\rho \sigma} (\delta_{k,0} -1 )  \comma \ee
true for all $\rho, \sigma, k,r,s, m,n$. Here $Rm$ is the result of the operation $R$ on the 
set $m$, so from (\ref{autoR}) it follows that  $(Rm)_j = m_{j-1} + \rho (1-\delta{j,1})$.

\subsubsection*{Other elementary relations}
There are no zeros or poles of $T_{pq}$ or $\tilde{T}_{pq}$ in the central domain ${\cal D}_0$, which 
is when $m = n = 0$, so
\be \label{phys}
e(0,0,k,0,0,|0,0) = 0 \period \ee

The model also possesses a reflection symmetry \cite{BPAY88}. Let $S$ be the automorphism 
defined in (\ref{RSU}), (\ref{autoS}). Then the Boltzmann weights satisfy $W_{Sq,Sp}(n) = W_{pq} (n)$, 
$\overline{W}_{Sq,Sp}(n) = \overline{W}_{pq} (-n)$. Replacing $p,q$ by $Sq,Sp$ therefore 
reflects the lattice about its SW-NE axis. This does not change the partition function, so
$T_{Sq,Sp} = T_{pq}$.

It also leaves $\Theta_{ij}$ unchanges, while replacing $\overline{\Theta}_{ij}$ by 
$\overline{\Theta}_{ji}$. The product over $i,j$ in (\ref{defG}) is symmetric in $i,j$, so $G_{pq}$ is unchanged
and from (\ref{defTT})
\be \tilde{T}_{Sq,Sp} \eq \tilde{T}_{pq} \period \ee
From (\ref{prodform}) it follows that
\be \label{symmS}
e(1-\sigma, 1-\rho,k,-s,-r|Sn,Sm ) = e(\rho,\sigma,k,r,s |m,n) \comma \ee
where $Sm$ is $S$ acting on the set $m$, so from  (\ref{autoS}) $(Sm)_j = -m_{N+1-j}$. 
Similarly for $Sn$.

\subsubsection*{A bilinear ansatz}

To give the free energy, the rotation and inversion relations must always be 
supplemented by analyticity assumptions, We know from the result (\ref{defVpq}) that 
$e(\rho,\sigma, k,r,s|m, n )$ is linear in the $m_j$ and $n_j$ separately, and 
hence bilinear in their combination. This seems to be a basic property that one may have expected, 
so we assume that there exist coefficients $A, B, C, D$ such that 
\bd
e(\rho,\sigma, k,r,s|m, n ) \eq 
A(\rho, \sigma, k, r, s) + \sum_{j=0}^{N-1} \left[ B(\rho, \sigma, k, r,s| j) m_j
+  C(\rho, \sigma, k, r,s| j) n_j \right] + \ed
\be  \label{ansatz}
\sum_{i=0}^{N-1}  \sum_{j=0}^{N-1}  D(\rho, \sigma, k, r,s|i, j) m_i n_j
 \ee
for all integers $m_j, n_j$. For generality we allow the coefficients to explicitly 
depend on the types $r, s$ of the Riemann sheet.

There are thus $4 N^3 (N+1)^2$ unknown coefficients to determine. Rather than proceed 
fully algebraically, we have looked at modest values of $N$ (not bigger than 12) and used 
Mathematica (with 
$m_j, n_j$ arbitrary) to determine whether the number of variables fixed by the various 
equations. Thus the following remarks are extrapolated conjectures from our observations.

We take $N \geq 3$. As we  noted above, the Ising case $N = 2$ is special in that 
then the $m_j$ and $n_j$ are not 
linearly independent, but satisfy $m_{1-\rho} = n_{1-\sigma} = 0$. All the solutions we write 
down are valid for $N = 2$, but there may be other solutions, so that our remarks concerning 
the number of solutions may not apply. 

First we substituted the ansatz (\ref{ansatz}) into the consistency 
condition  (\ref{cons}) and observed that the number of undetermined coefficients reduced to $2N^2(N+1)$.
Since one can fix $k, r-s$ and $(-1)^{\rho + \sigma}$ in this condition, this means that 
for each such case there are just $N+1$ unknown coefficients, i.e. $N+1$ linearly independent 
functions satisfying (\ref{cons}).

We know what these functions are. The functions $\tilde{T}_{p,U^{\alpha} q}, 
\tilde{T}_{U^{\alpha} p, q}, \tau_2 (p,U^{\alpha} q)$ are all meromorphic on the Riemann 
surface. Their exponents, given by (\ref{psidef}), (\ref{st2}), (\ref{st3}),
 must therefore satisfy (\ref{cons}). Taking 
$\alpha = 0, \ldots , N-1$, we obtain $3 N$ solutions of (\ref{cons}). They cannot all 
be linearly independent, but those in the  first set certainly are, so that gives us $N$ solutions.
The remaining one is a constant - independent of the $m_j$ and $n_j$.

It follows that $e(\rho,\sigma, k,r,s|m, n )$ must be of the form
\be e(\rho,\sigma, k,r,s|m, n ) = 
\gamma(\rho, \sigma, k, r-s) + \sum_{\alpha =0}^{N-1} 
c(\rho, \sigma, k, r-s |\alpha) \psi(\rho,\sigma,k,\alpha |m,n) \ee
where the coefficients $\gamma, c$ satisfy  
\bd
\gamma(\rho, \sigma, k, \lambda) \eq \gamma(1-\rho, 1-\sigma, k, k-\lambda)  \sep
c(\rho, \sigma, k, \lambda,\alpha) \eq c(1-\rho, 1-\sigma, k, k-\lambda,\alpha) \comma \ed
so there are $2 N^2 (N+1)$ independent coefficients $\gamma, c$, as yet undetermined.
The consistency condition (\ref{cons}) is now satisfied.

For $N$ odd, the rotation relation (\ref{rotnb}) reduces the number of undetermined 
coefficients to $N (N+1)$. Then the inversion relation (\ref{invnb}) fixes all the 
coefficients, giving the solution (\ref{defVpq}). The extra relations (\ref{phys}), 
(\ref{symmS}) are then satisfied.

For $N$ even and greater than 2, (\ref{rotnb}) reduces the number of undetermined 
coefficients  to $N(5N+6)/4$. Then  (\ref{invnb}) further reduces the number to $[(N+4)/4]$
(so for $N = $ 4, 6, 8, 10, 12 the numbers are  2, 2, 3, 3, 4).
The other relations (\ref{phys}), (\ref{symmS})  are satisfied by this solution, so do not 
reduce the number of undetermined 
coefficients  any further. The solution is then of the form
\bd \epsilon_{pq}(k) + h(\rho -\sigma -k +2 r - 2s) f(\rho, \sigma, k| m, n) \comma \ed

\noindent where $\epsilon_{pq}(k)$ is given by (\ref{defVpq}) and 
\bd
f(\rho, \sigma, k| m, n) \eq g_{k+1} - g_{k-1} - m_{k+\sigma} + n_{\rho-k} 
+ \sum_{\alpha = 0}^{N-1} [(-1)^{\alpha - k - \sigma } m_{\alpha} -
(-1)^{\alpha +k - \rho} n_{\alpha} ] \period \ed
These functions have the weak difference property that they are unchanged (for $N$ even) by 
incrementing all of $m_0, \ldots, m_{N-1}, n_0, \ldots, n_{N-1}$ by the same arbitrary integer, 
so no further restrictions can be obtained by making this requirement. 

\noindent The coefficients $h(j)$ are integers, subject only to the constraints
\bd h(j) = h(N-j) = h(N+j) \comma \; {\rm and} \; \; \; h(j) = 0 \; \; {\rm if } \; \; j \; \; 
{\rm is \; \; odd } \period \ed
It follows that just $[(N+4)/4]$ of them (all with $j$ even) remain undetermined.

To summarize: if $N$ is odd, the rotation and inversion relations, together with the consistency 
condition (\ref{cons}) and the bilinear ansatz (\ref{ansatz}), are sufficient to determine the 
exponents $e(\rho,\sigma, k,r,s|m, n )$. Surprisingly,  one does not need the 
analyticity and non-zeroedness of $T_{pq}$ in the central physical domain ${\cal D}_0$, i.e. 
the relation (\ref{phys}).

The same is true for $N$ even and $\rho-\sigma - k $ odd.
In both these cases we find that there is no explicit dependence of $e(\rho,\sigma, k,r,s|m, n )$ on the 
types $r,s$ of the Riemann sheet.

For $N$ even  and $\rho-\sigma - k $ even they are not quite sufficient, but they leave only
$[(N+4)/4]$ parameters to be determined. If one also (guided by the other cases) assumes that 
$e(\rho,\sigma, k,r,s|m, n )$ is not explicitly dependent on $r$ or $s$, then there is only one 
undetermined coefficient left.

As we remarked above, a knowledge of the orders of the poles and zeros of a meromorphic 
function does not
by itself fix the function, but it goes a long way towards it. For instance, one might 
observe that the orders $\epsilon_{pq}(k)$ satisfy the relations (\ref{exprelns}) and 
then guess the full set of recursion relations (\ref{Trcontp1}) - (\ref{2ndratio1}), 
and hence (\ref{Trcontp}) - (\ref{2ndratio}). One could then test these relations 
from series expansions. They imply the vital Assumption 2 of \cite{paper1}, so one 
could then
use the method of \cite{paper1}  to obtain the result (67) therein.

\section{Summary}

We have shown that the partition function per site $T_{pq}$ lives on a Riemann surface with an 
infinite number of sheets, each sheet corresponding to a point an $(2N-1)$-dimensional lattice. However, 
one should note that  adjacent sheets need not correspond to any 
simple geometric definition of adjacent points: 
the adjacency rules are given by (\ref{mnbrs}), (\ref{nnbrs}).

It is a meromorphic function on this surface, with zeros and poles only when $t_p = \omega^k t_q$
for $k = 0, \ldots , N-1$. The orders of these zeros (the negative of the order of the  poles) 
are bilinear in the integers $m_0, \ldots ,n_{N-1}$ that specify the sheet. If we assume this 
bilinearity (with coefficients that may depend on $k$ and the parities and types of the Riemann 
sheet), then for $N$ odd they can be obtained from the rotation and  inversion relations. For $N$ even
this procedure does not uniquely
fix the orders, but does determine them to within a small number $[(N+4)/4]$ of 
free parameters.

The integers $m_0, \ldots ,n_{N-1}$ (more precisely the differences of the $m_j$, 
and of the $n_j$) are connected with 
the variables of the hyperelliptic
paarametrization of the chiral Potts model \cite{Kyoto} (with $u, v$ therein interchanged). 
We hope to discuss this point in a later paper, and at least present an elliptic function
expression for $\tau_2(t_q)$ in the case $N = 3$.

A significant motivation for this work has been the still outstanding problem of
obtaining the spontaneous magnetization (order parameter) of the chiral Potts model. There is an elegant 
conjecture
(eqn 3.13 of  \cite{Howes83} , eqn. 1.20 of \cite{AMPT89} , eqn. 15
of \cite{HenkelLacki89}) for this property, which is almost certainly true, but has not been
proved. Following the method of Jimbo, Miwa and Nakayashiki \cite{JMN93}, the author 
has derived functional relations for a generalized order parameter \cite{Baxter1998a},  
\cite{Baxter1998c}. These have a 
similar structure
to the inversion/rotation relations for the free energy. If one could solve them, 
then  one would have verified the conjecture.

It was therefore the author's hope that the techniques of this paper could be applied to
solving the order  parameter relations. It must be admitted that preliminary results are not 
encouraging. If we assume that the function lives on the same Riemann surface as $T_{pq}$
(wwhich is not obvious), and that the orders (exponents) of the poles and zeros when 
$t_p^N = t_q^N$ are bilinear in the $m_j, n_j$, then for
$N$ odd we find the solution is unique, but is merely the ``wrong solution''
we obtained in equation 72 of \cite{Baxter1998a}. So we 
are no further forward.

It is possible that this function has poles and zeros other than those when 
$t_p^N = t_q^N$: we have found  some suggestion of this in preliminary low-temperature 
expansion calculations, and hope to discuss this in a later paper. If so, then
the order parameter function may be much more complicated than $T_{pq}$, and the simple 
ideas we used here may need considerable expansion.

\end{document}